\documentstyle[psfig]{mn2e}
\newcommand{\etal}{{et al.~}}
\newcommand{\lta}{\la}
\newcommand{\gta}{\ga}

\newcommand{\kms}{\>{\rm km}\,{\rm s}^{-1}}

\newcommand{\cm}{\>{\rm cm}}

\newcommand{\kpc}{\>{\rm kpc}}
\newcommand{\Msun}{\>{\rm M_{\odot}}}
\newcommand{\msun}{\>{\rm M_{\odot}}}

\newcommand{\beq}{\begin{equation}}
\newcommand{\eeq}{\end{equation}}

\newcommand{\erg}{\>{\rm erg}}

\newcommand{\apj}{ApJ}
\newcommand{\apjs}{ApJS}

\newcommand{\mnras}{MNRAS}
\newcommand{\aap}{A\&A}

\newdimen\hssize
\newdimen\hdsize 
\begin{document}
            

\title[Preheating by Previrialization]
      {Preheating by Previrialization and its Impact on Galaxy Formation}
\author[H.J. Mo et al.]
       {H.J. Mo$^{1}$, Xiaohu Yang$^{1}$,   
        Frank C. van den Bosch$^{2}$, Neal Katz$^{1}$
        \thanks{E-mail: hjmo@nova.astro.umass.edu}\\
      $^1$ Department of Astronomy, University of Massachusetts,
           Amherst MA 01003-9305, USA\\
      $^2$ Department of Physics, Swiss Federal Institute of
           Technology, ETH H\"onggerberg, CH-8093, Zurich,
           Switzerland}


\date{}


\maketitle

\label{firstpage}


\begin{abstract}
  We  use recent  observations of  the HI-mass  function  to constrain
  galaxy formation.   The data  conflicts  with the
  standard  model where most  of the  gas in  a low-mass  dark matter halo is
  assumed to settle into a disk  of cold gas that is depleted by star
  formation  and  supernova-driven  outflows  until the  disk  becomes
  gravitationally stable. Assuming a star formation threshold density
  supported by both theory and observations 
  this model predicts  HI masses  that are  much too  large. The reason
  is simple: supernova  feedback requires star formation, which
  in turn requires a high surface  density for the gas. Heating by the
  UV  background can  reduce the  amount of  cold gas  in  haloes with
  masses  $<10^{9.5}  h^{-1}{\rm  M}_\odot$,  but is  insufficient  to
  explain the  observed HI mass  function.  A consistent model  can be
  found if low-mass haloes are  embedded in a preheated medium, with a
  specific  gas entropy  $\sim 10\,{\rm  keV\,cm^{2}}$.   In addition,
  such  a model  simultaneously  matches the  faint-end  slope of  the
  galaxy  luminosity  function without  the  need  for any  supernova
  driven outflows.  We propose a  preheating model where the medium
  around  low-mass haloes  is  preheated by  gravitational pancaking.  
  Since gravitational tidal fields  suppress the formation of low-mass
  haloes while  promoting that of pancakes, the  formation of massive
  pancakes  precedes that  of the low-mass  haloes  within them.   We
  demonstrate that  the progenitors of present-day  dark matter haloes
  with $M\la  10^{12}h^{-1}\msun$ were embedded in  pancakes of masses
  $\sim 5\times  10^{12}h^{-1}\msun$ at  $z\sim 2$.  The  formation of
  such pancakes heats the gas to a temperature of $5\times 10^5{\rm
    K}$ and compresses it to an  overdensity of $\sim 10$.  Such gas has
  a cooling  time that exceeds the  age of the Universe  at $z\la 2$,
  and has  a specific entropy of $\sim  15\,{\rm keV\,cm^{2}}$, almost
  exactly  the amount  required to  explain  the stellar  and HI  mass
  functions.
\end{abstract}




\section{Introduction}

The  cold dark  matter (CDM)  model  of structure  formation has
proven a  very successful  paradigm for understanding  the large-scale
structure  of the  Universe. However,  as far  as galaxy  formation is
concerned, a number of  important issues still remain. A long-standing
problem is that  CDM models in general predict  too many low-mass dark
matter  haloes.  The  mass  function of  dark  matter haloes,  $n(M)$,
scales with halo mass roughly  as $n(M)\propto M^{-2}$ at the low-mass
end.  This is in strong contrast with the observed luminosity function
of galaxies, $\Phi (L)$, which has a rather shallow shape at the faint
end,  with $\Phi(L)  \propto  L^{-1}$.  To reconcile  these
observations with  the CDM paradigm, the efficiency  of star formation
must be a  strongly nonlinear function of halo  mass (e.g.  Kauffmann,
White \& Guiderdoni 1993; Benson  \etal 2003; Yang \etal 2003; van den
Bosch \etal 2003b).  One of the biggest challenges in galaxy formation
is  to  understand the  physical  origin  of  this strongly  nonlinear
relationship.

Another important but related  challenge  is to understand the baryonic 
mass fraction 
as a function  of halo mass.  The baryonic mass  in dark matter haloes
maybe roughly divided into three components: stars, cold gas, and hot
gas.  In the most naive picture  one would expect that each halo has a
baryonic mass fraction  that is close to the  universal value of $\sim
17$  percent\footnote{Corresponding   to  a  $\Lambda$CDM  concordance
  cosmology  with  $\Omega_B  h^2  = 0.024$  and  $\Omega_m  h^2=0.14$
  (Spergel \etal  2003)}. In this naive picture one can  achieve a low
ratio  of stellar  mass to  halo mass  by keeping  a  relatively large
fraction of the  baryonic mass hot, either by  preventing the gas from
cooling or  by providing a heating  source that turns  cold gas into
hot  gas.   Alternatively,  one  may  achieve  a  low  star  formation
efficiency in low-mass haloes by making the total baryonic mass
fraction lower in lower mass haloes, which can be achieved either by
blowing baryons  out haloes  or by preventing  baryons from ever
becoming bound to haloes in the first place.

At the  present it is unclear which of  these scenarios
dominates.   To  a  large  extent  this ignorance  owes  to  the  poor
observational  constraints on  the baryonic  inventory as  function of
halo mass.  Historically, most  observational studies of galaxies have
focussed on  the stellar  light. Although this  has given us  a fairly
detailed consensus of the relation between stellar mass (or light) and
halo mass  (e.g., Yang  \etal 2003, 2005;  van den Bosch  \etal 2003b;
Tinker \etal  2004), less information  is available regarding  the hot
and  cold gas  components.  Hot,  tenuous gas  in low  mass  haloes is
notoriously  difficult  to  detect,  making our  knowledge  of  the
relation between halo  mass and hot gas mass  quite limited.  Based
on X-ray observations of a few relatively massive spiral galaxies, the
gas mass in  a hot halo component appears to be  small (e.g. Benson et
al. 2000). For  cold gas the situation  has improved substantially
in recent years, owing to the completion of relatively large, blind 21-cm
surveys (e.g.   Schneider,  Spitzak  \& Rosenberg  1998;
Kraan-Korteweg \etal  1999; Rosenberg  \& Schneider 2002;  Zwaan \etal
2003,  2005).  Using  these surveys,  it  is now  possible  to
obtain  important   constraints  on  galaxy   formation  (see  Section
\ref{sec:coldgas}).

When modelling galaxy formation,  the process most often considered to
suppress star formation in low  mass haloes is feedback from supernova
explosions.   As shown  by Dekel  \& Silk  (1986) and  White  \& Frenk
(1992),  the total  amount of  energy  released by  supernovae can  be
significantly larger  than the binding energy  of the gas  in low mass
haloes. Therefore,  as long as  a sufficiently large fraction  of this
energy  can  be  converted  into  kinetic  energy  (often  termed  the
`feedback efficiency'),  one can in  principle expel large  amounts of
baryonic material  from low mass haloes, thus  reducing the efficiency
of  star   formation.   Indeed,  semi-analytical   models  for  galaxy
formation that  include a simple  model for this feedback  process are
able  to reproduce  the  observed slope  of  the faint-end  luminosity
function  in   the  standard  $\Lambda$CDM  model,   if  the  feedback
efficiency is taken to be  sufficiently high (e.g., Benson \etal 2003;
Kang \etal 2005).

An important question, however,  is whether such high efficiencies are
realistic. For example, detailed hydrodynamical  simulations by Mac-Low
\& Ferrara (1999) and Strickland \& Stevens  (2000) show that
supernova  feedback  is far  less  efficient  in  expelling mass  than
commonly assumed because the onset  of Rayleigh-Taylor instabilities
severely limits the mass loading  efficiency of galactic  winds.

This prompted investigations  into alternative mechanisms  to
lower  the  star  formation   efficiency  in  low  mass  haloes. Another
possibility is  that photoionization heating  by the UV  background may
prevent gas  from cooling into  low-mass haloes (e.g.,  Efstathiou 1992;
Thoul  \& Weinberg 1996) by increasing its temperature.
Numerical simulations have shown  that this effect is only efficient in 
dark matter haloes with $M\la 10^{10}h^{-1}\msun$ (e.g. Quinn et al. 1996, 
Gnedin 2000; Hoeft \etal 2005), since the gas is only heated to 
$\sim 10^4$ to $10^5$ K.
Although this might be sufficient to explain the relatively
low  abundance  of  satellite  galaxies  in  Milky  Way  sized  haloes
(Bullock,  Kravtsov \&  Weinberg 2000;  Tully \etal  2002), it  is 
insufficient  to explain  the faint-end  slope of  the  galaxy luminosity
function.  However, if  a different mechanism  were to heat the
intergalactic   medium  (hereafter   IGM)   to  even   higher
temperatures  (and  thus higher  entropies),  the  same mechanism  could
affect more massive haloes as  well.  Such a process is often referred
to as  `preheating'. Along  these lines, Mo  \& Mao  (2002) considered
galaxy  formation in  an IGM  that was  preheated to  high  entropy by
vigorous energy feedback associated with the formation of stars in old
ellipticals  and bulges and  with active  galactic nuclei (AGN)  activity at
redshifts  of 2 to  3.  They showed  that such a  mechanism can
produce  the entropy  excess observed  today in  low-mass  clusters of
galaxies  without destroying the  bulk of  the Ly-$\alpha$  forest. In
addition, it  would affect the  formation of galaxies  in low  mass haloes
whose  virial temperature is  lower than  that of  the preheated  IGM. 
Numerical   simulations   show  that   such   preheating  may   indeed
significantly lower  the gas mass  fraction in low mass  haloes (e.g.,
van den Bosch, Abel \& Hernquist 2003a; Lu et al.  in preparation).

In this paper  we investigate an alternative mechanism  for creating a
preheated  IGM. Rather  than  relying  on star  formation  or AGN,  we
consider the  possibility that the  collapse of pancakes  (also called
sheets) and  filaments heats  the gas  in these structures
and that the  low mass haloes  within them form  in a
preheated  medium.   Although  the  standard picture  of  hierarchical
formation  is  one  in  which  more  massive  structures  form  later,
gravitational tidal fields suppress  the formation of low-mass haloes,
while  promoting   the  formation  of   pancakes.   Consequently,  the
formation of massive pancakes will precede that of low-mass dark matter
haloes, which form within them.  In  this paper  we show  that the  shock
heating associated with pancake collapse at $z \lta 2$ can heat the associated
gas to sufficiently high entropy that the subsequent gas accretion into
the low mass haloes that form within these
pancakes is strongly  affected.  We demonstrate  that the
impact of this previrialization  is strong enough to explain both the
faint-end of  the galaxy luminosity function  and  the low mass
end of  the galaxy HI mass  function, without having to  rely on
unrealistically high efficiencies for supernova feedback.

The outline of the paper is as follows.  In \S~\ref{sec:coldgas},
we use  current observational results of  the HI gas  mass function to
constrain  star formation and  feedback in  low-mass haloes.   We show
that these  observations are difficult to reconcile  with the conventional
feedback model, but  that a consistent model can  easily be found if low-mass
haloes     are    embedded    in     a    preheated     medium.     In
\S~\ref{sec:preheating}  we describe  how shocks  associated with
the formation of pancakes can  preheat the gas around low-mass haloes. 
In \S~\ref{sec:GF}  we discuss our results in  light of existing numerical
simulations and discuss the impact of our results on the properties of
galaxies and  the intergalactic medium.   We summarize our  results in
\S~\ref{sec:concl}.

\section{The Formation of Disk Galaxies}
\label{sec:coldgas}

\subsection{Observational Constraints}

\begin{figure*}
\centerline{\psfig{figure=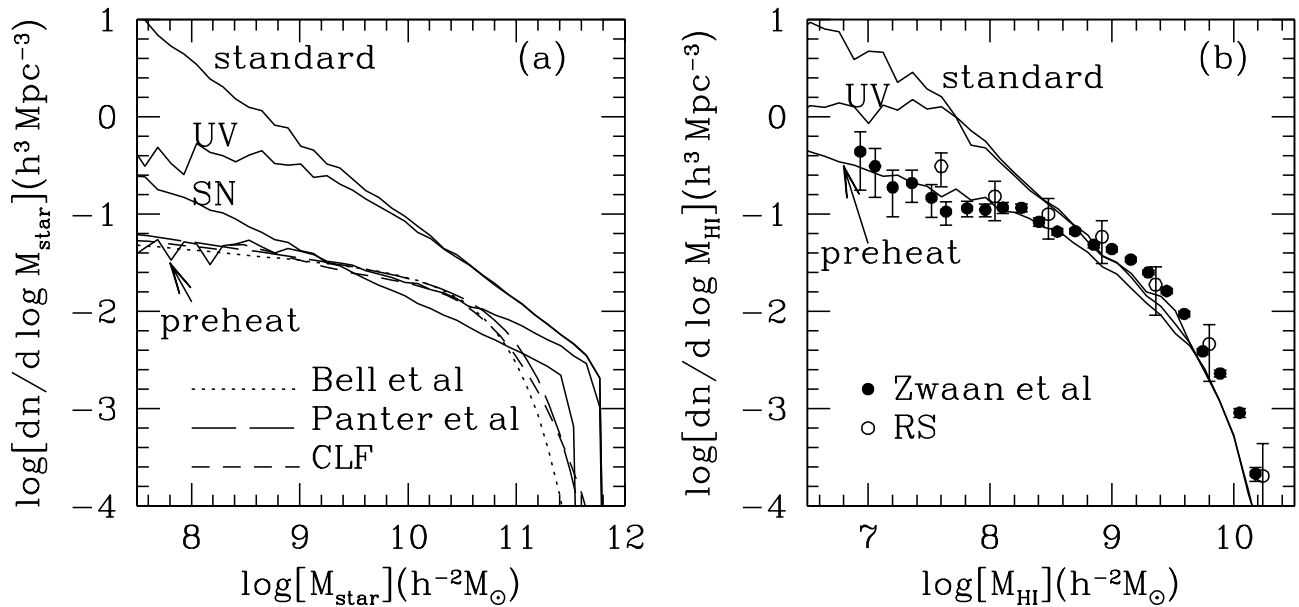,width=\hdsize}}
\caption{
  (a) The stellar mass functions  predicted by the standard model, the
  model with heating  by the UV background, and  the preheating model,
  as labelled. The curve labelled `SN' is the prediction of a model
  in which cold gas is heated by supernova explosions (see 
  text for details). 
  In addition, we  show the observational results of Bell
  \etal  (2003;  dotted curve)  and  Panter  \etal (2004;  long-dashed
  curve),  as well  as  the  stellar mass  function  of {\it  central}
  galaxies in dark matter haloes  derived from the CLF as described in
  the text  (short-dashed curve). (b) The  HI-mass functions predicted
  by the same three models  compared to the observed HI mass functions
  of  Rosenberg \&  Schneider  (2002; RS, open  circles)  and Zwaan  \etal
  (2005; solid dots).}
\label{fig:MF}
\end{figure*}

The models discussed below focus  on galaxies that form at the centers
of relatively low  mass haloes with $M \lta  10^{12} h^{-1} \Msun$.
To constrain  these models,  we first  derive the  stellar mass
function of central  galaxies using the conditional
luminosity  function  (CLF),  which  expresses how  many  galaxies  of
luminosity $L$ reside,  on average, in a halo of  mass $M$. Using both
the galaxy  luminosity function and  the luminosity dependence  of the
correlation length of  the galaxy-galaxy correlation function obtained
from the 2-degree Field  Galaxy Redshift Survey (2dFGRS, Colless \etal
2001), Yang \etal (2003) and van  den Bosch \etal (2003b) were able to
put tight  constraints on the CLF  (see also Yang \etal  2005 and van
den Bosch  \etal 2005).  As shown  in Yang \etal (2003),  the CLF also
allows one to  compute the average relation between  halo mass and the
luminosity of the  central galaxy (assumed to be  the brightest galaxy
in the halo).  We have determined this relation using the fiducial CLF
model considered in van den Bosch \etal (2005; Model 6 listed in their
Table~1).   To obtain  a stellar  mass function  for the central
galaxies, we  convert the 2dFGRS $b_J$-band luminosity  into a stellar
mass  using a  stellar mass-to-light  ratio $M_{\rm  star}/L =  4.0 \,
(L/L^\star)^{0.3}  \, (M/L)_\odot$  for $L  \leq L^\star$  and $M_{\rm
  star}/L=4.0 (M/L)_\odot$ for $L  > L^\star$, which matches the
mean relation  between stellar mass and  blue-band luminosity obtained
by  Kauffmann \etal (2003).   The resulting  stellar mass  function of
central galaxies is shown  in Fig.~\ref{fig:MF}(a) as the short-dashed
curve.  For  comparison, we  also plot the  stellar mass  functions of
{\it all}  galaxies obtained by  Bell \etal (2003) (dotted  curve) and
Panter, Heavens \& Jimenez (2004) (long-dashed curve). Given the large
uncertainties   involved  (see   Bell  \etal   2003  for   a  detailed
discussion),  these  stellar mass  functions  are  in remarkably  good
agreement  with each other, particularly for the low mass galaxies that are
the focus of this study.  The good agreement suggests that
most low  mass galaxies are  indeed central galaxies in  small haloes,
i.e. satellite galaxies do  not dominate  the stellar  mass function
(see also Cooray \& Milosavljevi\'c 2005).
   
In addition to the stellar mass function, we also constrain our models
using  the HI  mass function  of  galaxies.  With  large, blind  21-cm
surveys that  have recently been completed, this  HI-mass function has
now been estimated  quite accurately over a relatively  large range of
masses   (see  Zwaan   \etal   2005  and   references  therein).    In
Fig.\,\ref{fig:MF}(b), we  show the  recent results obtained  by Zwaan
\etal  (2005)  and  Rosenberg   \&  Schneider  (2002).   Both  HI-mass
functions are well fit by a Schechter (1976) function with a power-law
slope at the low mass end of  about $-1.3 \pm 0.1$.  Note that this is
slightly steeper than the power-law  slope at  the low  mass end  of the
stellar  mass function,  which is about $-1.16$ (Panter et al. 2004).  
Since  galaxy formation is  a process  that
involves  both stars  and  cold gas,  a  combination of  observational
constraints  on  the  luminosity  function and  the  HI-mass  function
provides  important constraints  on star  formation and  feedback.  In
fact, as  we show below, the  HI-mass function constraints 
are more generic,  and it is only by including them
we  are able to argue against  the standard supernova
feedback model.

\subsection{The standard model}
\label{ssec:standard}

In the standard picture of galaxy formation (White \& Rees 1978) it is
assumed that  gas cooling conserves  specific angular momentum.   As a
result, the baryons cool to form a centrifugally supported disk galaxy
(Fall \&  Efstathiou 1980).  In  what follows we investigate  the mass
functions of the cold gas and stars of disk galaxies that form within this
picture.  We make the simplifying
assumption  that each dark  matter halo  forms a  single disk  galaxy. 
Clearly this is a severe  oversimplification since it is known that
haloes, especially  more massive  ones, can contain more  than one
galaxy  and not every galaxy  is a disk galaxy.  However, we
are only interested in the properties  of low mass haloes, which to
good  approximation contain  a  single, dominant  disk  galaxy.   In
particular, the  studies of  van den Bosch  \etal (2003b),  Yang \etal
(2005), and Weinmann \etal (2005) show that in haloes with $M <
10^{12} h^{-1} \Msun$, the mass range considered here, the fraction of
late-type galaxies is  larger than 60 percent (see  also Berlind \etal
2003).  Thus, our simplified model will overpredict the number density
of disk galaxies in low mass haloes,  but not by more than a factor of
two.  Furthermore, we will conclude below that the main problem with the
standard model is one of
gas cooling, a problem that will exist independent of galaxy type.

To model the detailed structure of individual disk galaxies we use the
model of Mo, Mao \& White  (1998, hereafter MMW), which
matches a wide variety  of properties of  disk galaxies. 
Specifically, this  model assumes that  (i) the baryons have  the same
specific  angular  momentum  as  the  dark matter,  (ii)  they  
conserve  their specific angular  momentum when  they cool,  (iii) they
form  an exponential  disk,  and (iv)  the  halo responds  to the  gas
cooling  by  adiabatically  contracting.   Assumptions (i)  and  (iv)  are
supported by numerical simulations (van den Bosch \etal 2002; Jesseit,
Naab \&  Burkert 2002),  while assumption (ii)  is required  to obtain
disks of the  right size.  Finally, assumption (iii)  is equivalent to
assuming a particular distribution of specific angular momentum in the
proto-galaxy.  Haloes  are modelled as NFW spheres  (Navarro, Frenk \&
White 1997) with  a concentration that depends on  halo mass following
Bullock \etal (2001a),  and a halo spin parameter,  $\lambda$, that is
drawn from  a lognormal distribution with  a median of $\sim  0.04$ and a
dispersion of $\sim  0.5$ (e.g., Warren  \etal 1992; Cole \&  Lacey 1996;
Bullock \etal 2001b). The one free parameter in this model is the disk
mass fraction  $m_d$, defined  as the disk  mass divided by  the total
virial mass.   Since radiative cooling  is very efficient in  low mass
haloes  with  $M<10^{12}h^{-1}\msun$, we  start  our investigation  by
naively setting  $m_d$ equal to  the universal baryon  fraction, i.e.,
$m_d=0.17$.

In  a seminal  paper, Kennicutt  (1989) showed  that star  formation is
abruptly suppressed below a critical surface density.  This critical density
is close to that given by Toomre's stability criterion
\begin{equation}
\label{eq:Q}
\Sigma_{\rm crit}(R) = {\sigma_{\rm gas} \kappa(R) \over \pi G Q_{\rm crit}}\,,
\end{equation}
where  $\kappa(R)$ is the  epicyclic frequency,  $\sigma_{\rm  gas}$ is the
velocity  dispersion of  the  cold gas and $Q_{\rm crit}\sim 1$  (Toomre  1964).
This  critical density  determines  the fraction  of the gas  that  can form
stars (Quirk  1972).  Given  the  surface  density  of the  disk,
$\Sigma_{\rm disk}$, obtained using the MMW model described above, the
radius  $R_{\rm  SF}$ where the  density  of  the  disk  equals
$\Sigma_{\rm crit}$ can be calculated.  Following van den Bosch (2000)
we  assume that the  disk mass  inside this  radius with surface
density $\Sigma_{\rm disk} > \Sigma_{\rm crit}$ turns into stars,
i.e.
\begin{equation}
\label{mstar}
M_{\rm star} = 2 \pi\int_{0}^{R_{\rm SF}} \left[\Sigma_{\rm
    disk}(R) - \Sigma_{\rm crit}(R)\right] R\, {\rm d}R
\end{equation}
Kennicutt (1989) shows that $\sigma_{\rm gas} =  6 \kms$ and 
$Q_{\rm crit} \sim 1.5$
yields values of $R_{\rm
  SF}$ that correspond roughly to the radii where star formation is
truncated.  However, Hunter et al (1998) show that in low surface brightness 
galaxies $Q_{\rm crit} \sim 0.75$.  Therefore, to be conservative, we adopt
$\sigma_{\rm gas} =  6 \kms$ and $Q_{\rm crit} \sim 0.75$.
The  assumption that {\it  all} the gas  with $\Sigma_{\rm
  disk} > \Sigma_{\rm crit}$ has  formed stars is consistent with both
observations (Kennicutt 1989; Martin  \& Kennicutt 2001; Wong \& Blitz
2002;  Zasov \&  Smirnova  2005)  and with  predictions  based on  the
typical  star formation  rate in  disk galaxies  (Kennicutt  1998) and
their formation  times (see van  den Bosch 2001). We compute
the gas mass of each model galaxy using $M_{\rm gas} = \left(M_{\rm
    disk} - M_{\rm star}\right)$ where  $M_{\rm disk} = m_d M$. 
In other words, we assume that the gas surface density $\Sigma_{\rm gas}
= \Sigma_{\rm crit}$ inside $R_{\rm SF}$ and $\Sigma_{\rm gas} =
\Sigma_{\rm disk}$ outside $R_{\rm SF}$.
Finally, again to be conservative, we assume that the 
molecular hydrogen fraction is 1/2 
(e.g. Keres et al 2003; Boselli et al. 2002) 
so that the final HI mass of each galaxy, 
$M_{\rm HI} = 0.71 M_{\rm gas}/2$ where the factor of $0.71$ 
takes into  account the contribution of helium and
other heavier elements. 

Using  the halo mass  function given  by the  $\Lambda$CDM concordance
cosmology, and assuming that each halo  hosts a
single  disk  galaxy whose  properties  follow  from  $M$, $m_d$,  and
$\lambda$  as described  above,  we  obtain the  HI  and stellar  mass
functions    shown   as the  solid    curves   labeled    `standard'   in
Figure~\ref{fig:MF}. Since  we are  only interested in  low-mass haloes,
and since  our model does not  include any  processes that may
affect gas  assembly in massive  haloes, we artificially  truncate the
halo  mass function  at  $M =  5  \times 10^{12}  h^{-1} \Msun$, which
explains the abrupt  turnover of the model's stellar  mass function at high
masses. Not surprisingly,  our naive  model severely overpredicts  the stellar
mass  function, yielding an  abundance of  systems with  $M_{\rm star}
\simeq 10^8 h^{-2} \Msun$ that is  two orders of magnitude too large. 
In addition,  the model  predicts an HI  mass function at  $M_{\rm HI}
\lta 10^{8}h^{-2}\msun$ that  is more than 10 times  higher than
the  data.  Even  if we  reduce the number  density of  dark matter
haloes by  a factor  of 2,  to account for  the possibility  that some
isolated haloes may not host disk galaxies, the discrepancy remains a
factor of five.   Note that one might try to  lower the stellar masses
in our model by increasing $\Sigma_{\rm crit}$, but this leads to an
increase of the  HI masses, which are already  too large.  Similarly, a
decrease of  $\Sigma_{\rm crit}$  may improve the  fit of the  HI mass
function, but at the expense of  worsening the fit to the stellar mass
function. The failure of the standard model to
simultaneously fit the HI mass  function and the stellar mass function
is robust to the details  of star formation.  Fitting both
mass  functions  simultaneously  requires  either  a  modification  of the
cosmological parameters or additional physics to lower $m_d$. In what
follows, we consider both these possibilities separately.

\subsection{Cosmological Parameters}

One of  the main reasons that  the predicted HI mass  function is very
steep at the low-mass end is  that the halo mass function predicted by
the  standard  $\Lambda$CDM  model  is  also very steep at  the  low-mass
end.  Hence,  one way  to alleviate  the discrepancy  between  the model and
observations is to change the  cosmological parameters such that the low-mass
slope of the  halo mass  function becomes shallower.   Unfortunately, the
steep halo mass function is a very generic property of all CDM models.
In particular, the slope at  the low-mass end is almost independent of
cosmological parameters, including  the cosmic density parameter and the
amplitude of the primordial perturbations.  The only way to change the
slope of the  mass function at the low-mass end is  to assume that the
effective power index of  the primordial density perturbation spectrum
is significantly lower than  the scale invariant value.  However, such
models are not favored by the power spectrum derived directly from the
temperature fluctuations  of the  cosmic microwave background  and the
Lyman-$\alpha$ forest (e.g., Croft \etal 1999; Seljak \etal 2003), and
are  difficult,  if  not  impossible,  to reconcile  with  inflation.  
Another possibility  is that  the universe is  dominated by  warm dark
matter  (WDM) instead  of CDM,  so that  the power  spectrum  on small
scales is suppressed  by free stream damping of  the WDM particles. 
Here  again  observations  of  the Lyman-$\alpha$  forest  provide  a
stringent  constraint on  the particle  mass (Narayanan  \etal  2000). 
With particle masses in the  allowed range, the WDM  model yields a
halo mass  function that is  virtually indistinguishable from  that of
the CDM models in the halo mass range considered here. In other words,
within the parameter  space allowed by the data,  modifications of the
cosmological  parameters do  not have  any significant  impact  on the
results of the standard model presented above.

\subsection{Supernova Feedback}

Thus  far  we have  only  considered models  in  which  the disk  mass
fraction is  equal to  the universal baryon  fraction, i.e. where
$m_d=0.17$. We now consider  physical mechanisms that can lower $m_d$,
and investigate whether this allows a simultaneous match of the HI and
stellar mass functions. We first consider what has become the standard
mechanism,  namely feedback by  supernovae.  In  this model  each halo
acquires a baryonic mass fraction that is equal to the universal value
(0.17) but  $m_d$ is  reduced  since
supernovae inject large amounts of energy into the cold gas, causing it to
be ejected from the galaxy.

In semi-analytical  models of galaxy formation, two  schemes have been
proposed to  model this supernova feedback.  In  the `retention' model
considered  by Kauffmann  \etal  (1999), Cole  \etal (2000),  Springel
\etal (2001) and Kang \etal (2005), among others, part of the cold gas
in a  galaxy disk is  assumed to be  heated by supernovae to  the halo
virial temperature and is added to the hot halo gas for cooling in the
future.  Since  radiative cooling is  effective in galaxy  haloes, the
feedback efficiency  must be high enough to keep a  large amount of
the gas in the hot phase  (Benson \etal 2003).  However, if there is a
critical surface density below  which star formation ceases, no galaxy
disk  is  expected to  have  a  surface  density below  this  critical
density, because  otherwise the feedback from star  formation would be
insufficient to  keep most of the  gas hot.  
We plot an example of such a model in Figure~\ref{fig:MF}.  In this model we 
make the standard assumption that the rate at which cold gas 
is heated by supernova explosions is proportional to the star 
formation rate, ${\dot M}_{\rm reheat}=\beta {\dot M}_\star$, 
where $\beta=(V_{\rm hot}/V_c)^2$, with $V_c$ the circular velocity 
of the host halo, and $V_{\rm hot}$ an  ajustable parameter
(e.g. Benson et al. 2003). If the heated 
gas is not able to cool, the mass of the gas that can form stars 
will be $1/(1+\beta)$ times $m_d M-M_{\rm gas}$, where 
$M_{\rm gas}$ is the mass of the gas that remains in the disk. 
The curve labelled `SN' in Figure \ref{fig:MF} is the result
corresponding to $V_{\rm hot}=280\kms$. Although 
the stellar mass function is reduced significantly in this
model, the predicted slope at the low-mass end is much too steep.
Furthermore, the HI-mass function is unchanged
from the standard no-feedback model and so it is still unable to
match the observed HI-mass function.  In addition, this
model predicts fairly extensive haloes of hot, X-ray emitting gas,
which is inconsistent with observations (Benson \etal 2000).

An alternative  feedback model,  considered by Kauffmann  \etal (1999)
and Somerville \& Primack (1999), is the `ejection' model in
which the reheated gas is assumed  to be ejected from the current host
halo.  If the  initial velocity of the ejected gas  is not much larger
than the escape velocity of the  host halo, the gas will be recaptured
at a  later epoch as  the halo grows more massive by accreting  new material
from  its surroundings. Then , as in
the retention  model, the baryon fraction in the more massive halo will be
similar to the universal value, except that  a (typically  small) delay
time is added. Consequently, this model also cannot produce  disks  with  gas
surface densities below  the critical  density, which again
results in  a severe overproduction  of low HI mass systems.
Only if the supernova explosion energy
heats  the cold gas  to an energy  much greater than  the binding
energy  of  the  halo  can the  wind  escape  the  halo forever and potentially
reduce the number of low HI mass systems.
However,  there are  several problems  with this
scenario.   First,  as shown  by Martin  (1999) and  Heckman \etal
(2000), the  observed mass outflow rate in  starburst galaxies, which
are extreme  systems, is about twice the star  formation rate. 
This implies that  one can never achieve a disk  mass fraction that is
lower  than about  1/3 the  universal baryon  fraction, which  is not nearly
enough  to match the HI observations.  Second,  the numerical
simulations  of Mac-Low \&  Ferrara (1999)  and Strickland  \& Stevens
(2000) have shown that the mass  loss rates in quiescent disk galaxies
are much lower than those observed in starburst galaxies.  Third, as
shown  in Benson  \etal  (2003), the  feedback  efficiencies that  are
required to permanently eject  the gas are completely unphysical.  
Finally, as  shown in van  den Bosch (2002),  even if one  ignores all
these problems  and simply ejects the  gas forever, the  presence of a
star formation threshold density  still assures that the final surface
density of the gas is similar  to the critical surface density.  As is
evident  from Figure 5 in  that  paper, supernova  feedback that is
modelled with permanent ejection has  a drastic impact on the stellar
masses but leaves the gas mass basically unchanged.  Again, this owes
to the fact that gas ejection requires supernovae, which in turn requires star
formation,  which requires  a  gas surface  density  
that exceeds  the
critical   density.   Note  that   although  supernova   feedback  may
temporarily deplete the gas  surface density below the critical value,
the  ongoing cooling  of  new and  previously  expelled material  will
continue to  increase the  cold gas surface  density until  it exceeds
$\Sigma_{\rm crit}$, initiating  a new episode of star  formation and
its associated feedback.
As shown  in  van den  Bosch  (2002), this  results in  a
population of disk galaxies whose cold gas surface densities are, in a
statistical sense, similar to $\Sigma_{\rm crit}$.

In summary, although  supernova feedback may be tuned  to yield a good
match to the  low-mass end of the stellar mass  function, it has three
fundamental  problems: (i)  the efficiency  needed  seems unphysically
high compared  to what it achieved in  detailed numerical simulations,
(ii)  unless the hot  gas is  expelled from  the halo  indefinitely it
predicts haloes of hot, X-ray emitting gas which are inconsistent with
observations, and (iii) as demonstrated  here, if one takes account of
a  star formation threshold  density, which  is strongly  supported by
both theory and observations, it overpredicts the abundance of systems
with low HI  masses by almost an order of  magnitude.  
In short, the problem of matching the observed HI mass functions is one of
preventing gas from entering the galaxy in the first place. Standard
feedback schemes fail because even if all the gas is temporarily removed,
e.g. by a massive supernova outflow, it will just reaccrete more gas.
This is also why our arguments, although framed around disk galaxies,
should hold for all galaxies.
It is, therefore,
important to seek other solutions that are physically more plausible.

\subsection{Reionization and Preheating}

In the models considered above we assume that the  IGM accreted by the
dark matter haloes is cold, allowing all of the gas originally associated with
the halo  to be accreted eventually.  However, if  the gas in the  IGM is
preheated  to a  specific  entropy that  is comparable  to 
or larger than that
generated by the accretion shocks associated with the formation of the
haloes, not all  of this gas will be accreted into the  halo (e.g.  Mo \&
Mao 2002;  Oh \&  Benson 2003;  van den Bosch  \etal 2003a).   In that
case, some disks may start with  a gas surface density already
below the critical density, making their HI gas masses smaller.
Note that  this circumvents  the problem  with the
supernova feedback  scenario that requires star formation  and its inherent
high gas surface densities.

Let us  first consider photoionization  heating by the UV  background. 
After  reionization,  the  UV  background   can  heat  the  IGM  to  a
temperature of roughly $20,000$K.   As first pointed out by Blumenthal
\etal (1984),  such heating of the  IGM can affect  gas accretion into
dark matter haloes of low masses (see also Rees 1986; Efstathiou 1992;
Quinn et all 1996; Thoul \& Weinberg 1996).   Recent simulations
(e.g., Gnedin  2000; Hoeft
\etal 2005)  follow the detailed  evolution of the  UV background and
show that  at the  present time the  fraction of  gas that  can be
accreted into a dark matter halo of mass $M$ can be written approximately as
\begin{equation}\label{fB:UV}
m_{\rm gas}={f_{B}\over (1+M_c/M)^\alpha}\,,
\end{equation}
where $f_B$  is the universal baryon fraction.   Following Hoeft \etal
(2005), we  consider a model with  $M_c=1.7\times 10^{9} h^{-1}\msun$,
and $\alpha=3$.  Using  the same disk formation model  as described in
Section~\ref{ssec:standard},  and  assuming  that disks  with  surface
densities  below $\Sigma_{\rm  crit}$ do  not form  any stars,  we can
predict the cold  gas and stellar masses.
The resulting  stellar and  HI-mass functions are shown as
the  solid   lines  labelled  `UV'   in  Figure~\ref{fig:MF}.   Although
preheating by the  UV background clearly reduces both  the stellar and
HI-mass  functions  at  the  low-mass  end, the  strength of the
effect  fails to reconcile the models with
the data (see also Benson \etal 2002).

However, since  the predicted
cold gas mass is significantly reduced in haloes with masses below the
characteristic  mass scale  $M_c$,  this  motivated us  to  consider a
model  in which  the  IGM around  low-mass  haloes is  preheated to  a
temperature that is significantly higher than $20,000$K.  A higher
temperature corresponds
to a  larger $M_c$. As shown in  Lu \& Mo (2005,  in preparation), the
mass fraction of baryons that are  accreted by dark matter haloes in a
strongly preheated medium  is well described by equation~(\ref{fB:UV})
with $\alpha  \sim 1$. Therefore, as a test of this idea we  adopt $\alpha=1$
and choose $M_c = 5\times 10^{11}h^{-1}\msun$, which corresponds to an
initial   specific  entropy   for  the   preheated  IGM   of  $s\equiv
T/n_e^{2/3}\sim 10 {\rm kev\, cm^2}$ (where $n_e$ is the number density of
electrons). To determine the relationship between  $M_c$ and $s$ we assume that
$T$ is the virial temperature of a halo with mass $M_c$ at
the present time and that $n_e$ is the mean density of electrons within 
the halo. In  \S\ref{sec:preheating},  we
propose a  model that explains how  the IGM is preheated to
such a  level. Here  we examine how  this preheating affects  both the
HI-mass function and the stellar mass function

The  solid curves  labelled  `preheat' in  Figure~\ref{fig:MF} show  the
stellar and HI-mass functions predicted  by this model, using the same
disk  formation model described above.   Contrary to  the standard
model  and  the  reionization  model,  this  preheating  model 
agrees  with the  data fairly well for the low mass haloes that concern us here.
We underpredict the HI mass function at higher masses but remember to be
conservative we tried to make the HI mass function as small as possible.
For example, we took a small value $Q_{\rm crit}=0.75$, which is appropriate
for dwarf galaxies.  If we took $Q_{\rm crit}\sim 1.5$, which is 
appropriate for larger galaxies, it would increase the masses 
and make them better match the observations.
\begin{figure}
\centerline{\psfig{figure=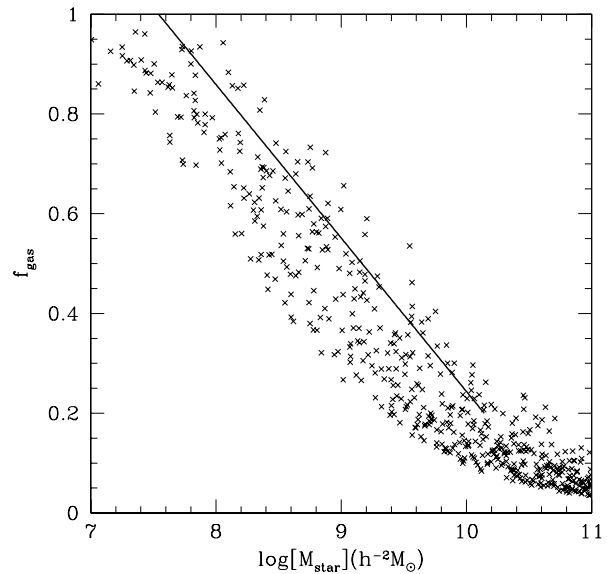,width=\hssize}}
\caption{The cold gas mass fraction, defined as the ratio 
  between the  mass of cold gas and  the total mass of  stars and cold
  gas,  as a  function of  stellar  mass predicted  by the  preheating
  model (crosses). The thick solid line  shows the observed mean 
  relation given
  by McGaugh \& de Blok (1997).}
\label{fig:fg}
\end{figure}

Unlike the supernova feedback model,  preheating can  simultaneously match
the HI  and stellar mass  functions. However, this does  not guarantee
that the  model also predicts  the correct ratio  of cold gas  mass to
stellar mass in individual galaxies. To test this, we compute for each
model galaxy  the cold gas  mass fraction, $f_{\rm gas}  \equiv M_{\rm
  gas}/(M_{\rm gas} + M_{\rm star})$.  Figure~\ref{fig:fg} plots $f_{\rm
  gas}$ as a  function of the stellar mass. The scatter
in the model predictions results from the scatter in the halo spin
parameters, and is comparable to the observed scatter (McGaugh \& de
Blok 1997).  The preheating model  predicts that the gas mass fraction
decreases   with   stellar  mass,   in   qualitative  agreement   with
the observations (McGaugh \& de Blok 1997; Garnett 2002). As already
demonstrated  in  van den  Bosch  (2000),  this  success is a direct consequence
of implementing a critical surface density for star formation.  Note
that the  model predicts  gas fractions that  are slightly  lower than those
observed. However, given the  uncertainties involved, both in the data
and in the model, and given the relatively large amount of scatter, we
do not consider this a serious shortcoming.

As  a  final  test  of  the  preheating  model  we  consider  gas
metallicities.   The  higher  gas  mass fractions  in  smaller  haloes
implies that  the metallicity of the  cold gas must be  lower in lower
mass  systems, which  is consistent  with observations  (Garnett
2002).  However, observations also show that the {\it effective} metal
yield decreases  with galaxy luminosity (Garnett 2002; Tremonti
\etal 2004), suggesting that some  metals generated by stars must have
been ejected from the galaxies and that the ejected fraction is larger
for fainter  galaxies.  This may  seem problematic for  the preheating
model considered here,  since we require no outflows to
match the  HI and  the stellar mass  functions.  However,
this does not exclude the possibility that significant amounts of {\it
  metals} have been  ejected from low-mass galaxies.  In  fact, as the
numerical  simulations  of  Mac-Low  \&  Ferrara  (1999)  demonstrate,
supernova feedback in quiescent disk galaxies is far more efficient at
ejecting metals than mass.  The  reason for this is that the metals are largely
produced by  the supernovae themselves, so  they are  part of the
hot bubbles of tenuous gas that make up the galactic winds. When these
bubbles  rupture owing to Raleigh-Taylor  instabilities  this
strongly metal enriched material, which has relatively little mass, is
blown away from  the disk by its own  pressure. Thus, although clearly
more  work is  needed to  test  this in  detail, we  believe that  the
observed effective metal  yields are  not at odds with  the preheating
model.

\section{Preheating by gravitational pancaking}
\label{sec:preheating}

\begin{figure*}
\centerline{\psfig{figure=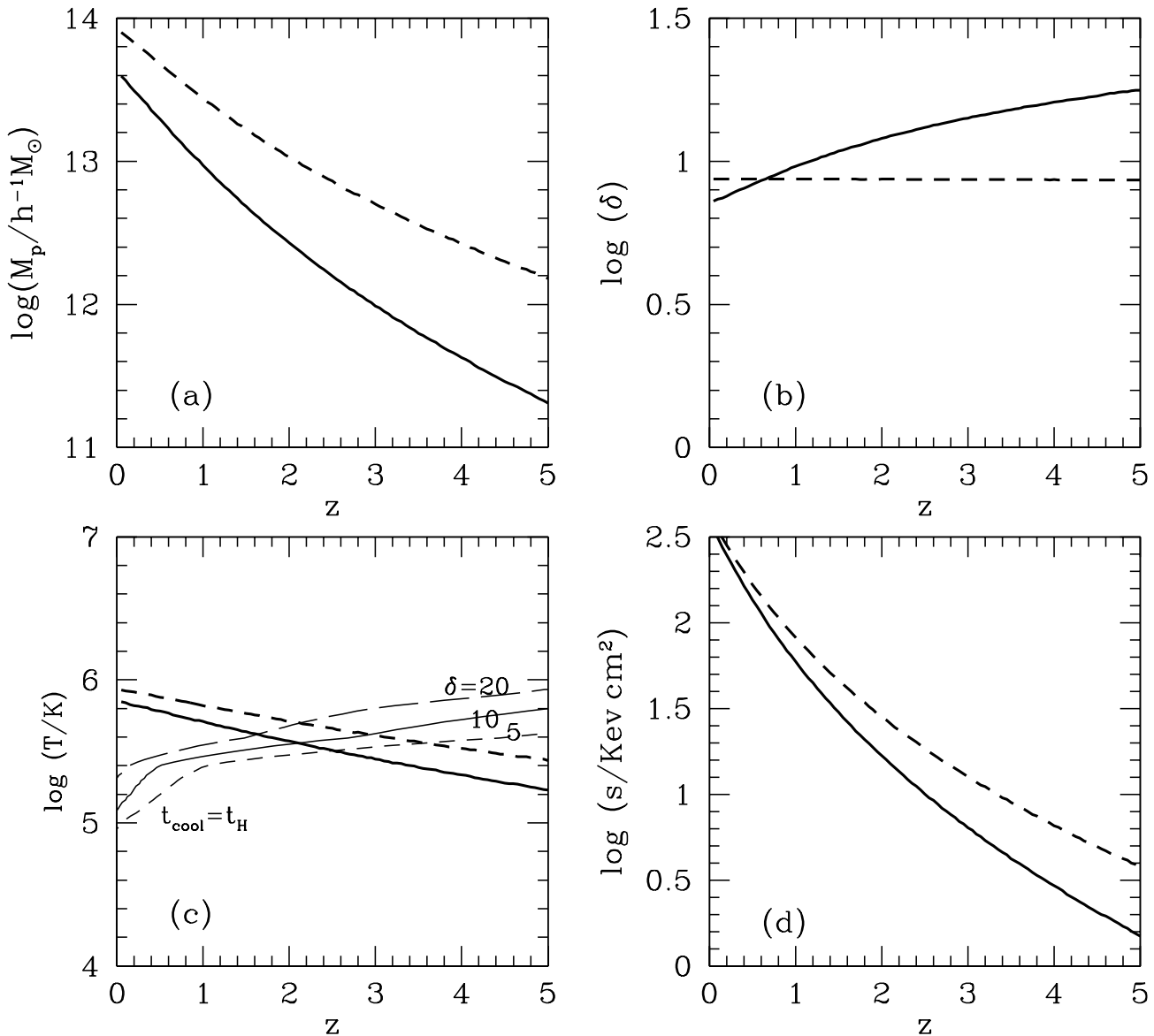,width=\hdsize}}
\caption{The mass [panel (a)],  overdensity [panel (b)], 
  gas temperature [panel (c)], and gas specific entropy [panel (d)] of
  pancakes around low-mass haloes at  the time of first axis collapse. 
  Thick solid curves assume  that perturbations around low-mass haloes
  have $e$ values  equal to the most probable  values corresponding to
  the mass scale in consideration,  while thick dashed curves show the
  results in which $e$ is  assumed to be a constant, i.e. independent
  of pancake  mass.  The  thin lines  in panel (c)  show the  loci of
  $t_{\rm  cool}  = t_{\rm  H}$  for  the three  indicated values  of  the
  overdensity.}
\label{fig:entropy}
\end{figure*}

In  the previous section  we have  shown that  a preheating  model, in
which the gas surrounding present-day  low mass haloes is preheated to
a specific entropy of $s\sim 10\,{\rm kev\,\cm^2}$, can simultaneously
match the low  mass ends of both the HI-mass function  and the stellar mass
function. Here we propose a  physical process that can
cause  such  preheating.  

We base our proposed model on the following  considerations.  In the
basic picture  of structure formation  in CDM cosmogonies, as
the universe expands, larger and larger objects collapse 
owing to  gravitational  instability.   The collapse  is
generically aspherical (Zel'dovich 1970), first  forming sheet-like pancakes
(first axis collapse), followed  by filamentary structures  
(second axis collapse),
and eventually  virialized dark matter  haloes (third axis  collapse). 
Thus, according to the ellipsoidal  collapse model, the formation of a
virialized halo requires the collapse  of all three axes (e.g. Bond \&
Myers 1996; Sheth, Mo \& White 2001). However, owing to the large scale tidal
field,  the density  threshold for  the formation  of a  low-mass halo,
i.e. one with a mass  below the characteristic mass, $M_*$,  defined as the
mass  at which  the rms  fluctuation is  equal to  unity, can  be much
higher than  that in the spherical collapse  model.  This delays the
formation of low  mass haloes relative  to the prediction
of the spherical  collapse model.  Conversely, the tidal field
accelerates the collapse of the first  (shortest) axis and hence
the formation  of a pancake can require  a density threshold 
much lower  than that for  spherical collapse.  Consequently,  many of
today's  low  mass  haloes,   i.e.   those  with  $M  \ll  M_*(z=0)\sim
10^{13}h^{-1}\msun$, formed in pancakes of  larger mass, which
formed before the haloes themselves.

In  the process  of pancake  formation,  the gas  associated with  the
pancake is  shock heated.  If the  temperature of the  shocked gas is
sufficiently high,  and if  the gas is  not able  to cool in  a Hubble
time, the  haloes embedded  in the pancakes  will have to  accrete their gas
from a preheated  medium, which as we showed in  the previous section may
have important implications for the formation of galaxies within those
haloes.   To  see  if  this   process  of  ``gravitational
pancaking'' (or previrialization, Peebles 1990)
can generate a preheated medium with the required specific
entropy,  we need  to examine  the properties  of the  pancakes within
which present-day low mass haloes  formed, and to understand how
the gas associated with these pancakes was shock heated.
 
To study the  first problem it is important to  realize that the bias
parameters of haloes  with $M\la 0.1 M_*$ have  similar values (Mo \&
White 1996; 2002; Jing 1999; Sheth \& Tormen 1999; Sheth, Mo \& Tormen
2001; Seljak \& Warren 2004;  Tinker \etal 2004).  This means that all
such  haloes  are,  in   a  statistical  sense,  embedded  within  similar
large-scale environments. At $z=0$, $M_\star\sim 10^{13} h^{-1}\msun$,
and  so all  haloes with  $M\la 10^{12}  h^{-1}\msun$ are  embedded within
similar environments.

According to the ellipsoidal collapse  model, the collapse of a region
on some  mass scale $M$  in the cosmic  density field is  specified by
$\delta$, the average overdensity  of the region in consideration, and
by $e$ and  $p$, which express the ellipticity  and prolateness of the
tidal shear  field in the neighborhood of that  region.
According to Sheth  \etal (2001),
the density threshold for the  formation of a virialized halo is given
by solving
\begin{equation}
{\delta_{\rm ec}(e,p)\over\delta_{\rm sc}} = 
1 + \beta\left[5(e^2\pm p^2) {\delta_{\rm ec}^2(e,p) \over
\delta_{\rm sc}^2} \right]^\gamma\,,
\end{equation} 
where  $\beta=0.47$,  $\gamma=0.615$,  and  $\delta_{\rm sc}$  is  the
critical overdensity  for spherical collapse.  For  a Gaussian density
field, the joint distribution of $e$ and $p$ for given a $\delta$ is
\begin{equation}
g(e,p\vert\delta) = {1125\over\sqrt{10\pi}} e\left(e^2-p^2\right)
\left({\delta\over\sigma}\right)^5 
e^{-5\delta^2(3e^2+p^2)/2\sigma^2}\,,
\end{equation}
where $\sigma$ is the rms fluctuation of the density field on the mass
scale  in  consideration  (Doroshkevich  1970).   For  all  $e$,  this
distribution peaks at $p=0$, and the maximum occurs at
\begin{equation}
\label{e_m}
e_{\rm max}(p=0\vert\delta) = {\sigma \over \sqrt{5} \delta}\,.
\end{equation}
Thus, the  most probable  value of  $e$ is related  to the  mass scale
through  $\sigma$. Using  this  relation, one  can  obtain a  relation
between the density threshold for collapse and the halo mass:
\begin{equation}
\delta_{\rm ec}(M,z) = \delta_{\rm sc}(z) \left\{1+0.47
\left[{\sigma^2\over \delta_{\rm sc}^2(z)}\right]^{0.615} \right\}\,
\end{equation}
(Sheth \etal 2001). 

The ellipsoidal  collapse model can  also be used  to determine  the density
threshold   for  the   formation  of   pancakes.   Based   on  similar
considerations,  one can  obtain a  corresponding relation  between the
collapse density threshold and the mass of the pancake:
\begin{equation}
\delta_{\rm ec}(M, z) = \delta_{\rm sc}(z)
\left\{1-0.56 \left[{\sigma^2\over \delta_{\rm sc}^2(z)}\right]^{0.55}
\right\}\,
\end{equation} 
(Shen, Abel,  Mo \& Sheth, in  preparation).  As one can  see, for low
peaks (i.e. $\delta_{\rm sc}/\sigma\ll  1$), the two thresholds can be
very  different, while  for high  peaks they  are similar.  Thus, the
effect of pancaking on subsequent halo formation is more important for
lower peaks, i.e.  for lower mass haloes with later formation times.

Given the above properties of  the collapse thresholds, we are able to
address the following question:  for low-mass haloes identified at the
present time, what is the nature of the pancakes within which their progenitor
haloes were  embedded at  an  earlier  time?  To  quantify  the formation  of
pancakes around a given halo rigorously, one needs to calculate
the  conditional  probability distribution for  the  overdensity of 
density  perturbations  on  various  scales  around  the  halo  and  the
corresponding tidal shear fields. It is beyond the scope of this paper
to   present  such a  detailed   analysis  here.    Instead,  we   use  the
cross-correlation  between  haloes and  the  linear  density field  to
determine the  average linear overdensity  around dark matter  haloes on
different scales.   We then use this  overdensity to characterize  the mean
environment of haloes of a given mass at the present time at earlier times.

According to the halo bias model (Mo \& White 1996), the average linear
overdensity within a  radius $r$ of a halo of mass  $M$ can be written
as
\begin{equation}
\label{bias}
{\overline \delta}_l(r) = b(M) {\overline\xi}_{\rm m}(r)\,,
\end{equation}
where $b(M)$ is the bias parameter for haloes of mass $M$, and
the average two-point correlation function of the linear density field
\begin{equation}
\label{averxi}
{\overline\xi}_{\rm m}(r) \equiv {3\over r^3}
\int_0^r \xi_{\rm m}(r') r'^2 {\rm d}r'\,,
\end{equation}
where  $\xi_{\rm m}(r)$ is the two-point correlation function.
The radius $r$  corresponds to a mass scale $M_r= (4\pi
r^3/3)  {\overline\rho}_0$,  where  ${\overline\rho}_0$  is  the  mean
density  of the universe  at $z=0$.   As we mentioned above,  the bias
parameter for present day haloes  with $M \la 10^{12} h^{-1}\msun$ is
independent of halo  mass, with  $b\sim 0.65$  (Sheth \etal
2001; Jing 1999;  Seljak \& Warran 2004; Tinker  \etal 2004). 
We adopt  this value to calculate ${\overline\delta}_l$.
It is then straightforward to calculate the
linear overdensity ${\overline \delta}_l$ as  a function of $r$ and
the corresponding  mass scale $M_r$.  If this  overdensity reaches the
overdensity  threshold for pancake  formation at  a given  redshift, a
pancake  of mass  $M_r$ will  form.   Since the
overdensity threshold  depends on the values  of $p$ and  $e$, in principle one
has to  calculate the joint distribution  function of $e$  and $p$ for
the appropriate mass  and overdensity, under the  condition that
the region  contains a halo  of mass $M$  at the present time.   As an
approximation,  we  assume that  both  $e$  and  $p$ take  their  most
probable values on the mass  scale in question, so that $p=0$ and
$e=e_{\rm  max}$.  We expect this  approximation  to  be valid  for
$M_r\gg M$, where  the correlation between the properties  of the halo
and its environment becomes weak (e.g. Bardeen et al. 1986).

For  a given  ${\overline\delta}_l$, $M_r$,  $p$  and  $e$,  we use  the
ellipsoidal  collapse model described in  Bond  \&  Myers (1996)  to follow  the
collapse of  a uniform ellipsoid embedded in  an expanding background
along  all  three  axes,   taking  into account  both  the  ellipsoid's
self-gravity and the external tidal field. Following Bond \& Myers, we
assume  that each  axis  freezes out  at  a constant  fraction of  its
initial radius, so  that the mean overdensity of  the collapsed object
is just the same as that in the spherical collapse model (see Bond \& Myers
1996  for  details).   The  solid curve  in  Fig.~\ref{fig:entropy}(a)
shows the mass  of the pancake that  forms
around present day low-mass haloes, as a function of $z$. Remember that at
earlier times these haloes have smaller masses. Also,
owing to the roughly constant bias parameter,
these results are almost independent of halo mass for
$M  \lta 10^{12} h^{-1} \Msun$.  As one  can see, the pancake
mass  decreases with  redshift, because  the overdensity  threshold
for collapse  is higher at higher $z$. At $z\sim 2$,  the pancake mass is
about     $10^{12.5}h^{-1}\msun$.       The     solid     curve     in
Fig.~\ref{fig:entropy}(b) shows the overdensity  of the pancake at the
time of formation.  This overdensity  increases with $z$, and is about
10 for  $z=1$ -  $2$.  The ellipsoidal  collapse model also  determines the
velocity  along the  first  axis at  the  time of  pancake formation.  
The gas associated with  the pancake will be
shocked.   If we  assume an adiabatic equation of state and that  the shock
is strong,  all the kinetic energy is transformed into internal energy.
We can calculate the post shock gas temperature as
\begin{equation}
T={3\mu V_1^2\over 16 k_B}\,,
\end{equation}
where  $k_B$ is  Boltzmann's  constant, $\mu$  is  the mean  molecular
weight, and $V_1$ is the velocity  along the first axis at the time of
shell crossing.  We plot this  temperature, as a function of  $z$, as
the thick  solid curve in  Fig.~\ref{fig:entropy}(c).  The temperature
decreases with  increasing redshift because the mass  of the pancake,
$M_p$, is smaller at  higher $z$ and $V_1  \sim H_0 R_p
\propto M_p^{1/3}$  where $R_p$ is the  Lagrangian radius of the  pancake. 
At $z\sim  2$, the temperature  is about $10^{5.5}{\rm  K}$.  Assuming
that the  gas overdensity is the same  as that of the  dark matter, we
can estimate the specific entropy generated in the shock as
\begin{eqnarray}
s = {T\over n_e^{2/3}} & = & 17 \times 
\left({\Omega_{\rm B,0}h^2\over 0.024}\right)^{-2/3}
\left({T\over 10^{5.5}{\rm K}}\right)\nonumber\\
 & & \times \left({1+\delta\over 10}\right)^{-2/3}
\left({1+z\over 3}\right)^{-2} {\rm keV}\cm^2\,,
\end{eqnarray}
where we have taken $\mu=0.6$, valid for a completely ionized 
medium dominated by hydrogen and helium. 
As shown by the solid
curve  in  Fig.\,\ref{fig:entropy}(d),   $s$  increases  rapidly  with
decreasing redshift,  mainly owing to  the decreasing gas  density.
At  $z  \sim  2$  the  resulting entropy  is  $s  \sim  15  {\rm
  KeV}\cm^{2}$. For  $z \lta 2.5$,  pancake formation results  in a
preheated IGM with $s \gta  10 {\rm KeV}\cm^{2}$, which corresponds to
the value  adopted in the  preheating model discussed in  the previous
section.

The results presented here are  based on the assumption that both $e$
and $p$ (which specify the local tidal field) take their most probable
values.   In reality, the  tidal field  around a  point in  a Gaussian
density field must be coherent over a finite scale. Since the low-mass
haloes at $z=0$ are low peaks,  typically with large values of $e$, it
is possible that the value of $e$ for the region around such a halo is
larger  than  the most  probable  value.  
Without  going into  detailed calculations that include
correlations of the  tidal field on different scales, we
consider an extreme case in which we set $e=0.45$ for all $M_r$.  This
value  for  $e$  is  approximately  equal  to that  for  a  peak  with
$\delta/\sigma=1$.   The thick  dashed curves  in the  four  panels of
Fig.~\ref{fig:entropy} show the results for this extreme model. 
Because the assumed value of $e$ is larger than the most probable
value, a  given  mass pancake forms earlier and correspondingly
the overdensity  of the pancake is  lower. However, for  $z\la 3$, the
change in $T$ is  less than $50\%$ and the change of  $s$ is less than a
factor of two.

The  preheating  by  gravitational   pancaking  is  expected  to  have
important consequences  for the subsequent accretion of  gas into dark
matter haloes  only if  the heated gas  cannot cool  efficiently.  The
cooling time of the heated gas can be written as
\begin{eqnarray}
t_{\rm cool} & \sim & 6.3 \, \Lambda_{-23}^{-1} \,
\left({\Omega_{\rm B,0} h^2\over 0.024}\right)^{-1}
\left({T\over 10^{5.5}{\rm K}}\right) \nonumber\\
 & & \times \left( {1+\delta\over 10} \right)^{-1}
\left({1+z\over 3}\right)^{-3} \,{\rm Gyr}\,,
\end{eqnarray}
where $\Lambda_{-23}$  is the cooling function in units  of $10^{-23} \erg
\sec^{-1}  \cm^3$.   This time  should  be  compared  with the  Hubble
time,\footnote{Our  notation  is such  that  ${\cal  H}(z)=1$ for  an
  Einstein-de  Sitter   Universe,  which   is  also  valid  to  good
  approximation for the $\Lambda$CDM concordance cosmology at $z \gta
  1$.}
\begin{equation}
t_{\rm H} = 5.0 \left({h\over 0.7}\right)^{-1}
\left({\Omega_0\over 0.3}\right)^{-1/2}
\left({1+z\over 3}\right)^{-3/2}
{\cal H}(z)\,{\rm Gyr}\,,
\end{equation}
where ${\cal  H}(z)=\Omega_0^{1/2} (1+z)^{3/2} H_0/H(z)$  and $H(z)$ is
Hubble's constant at redshift $z$. In Fig.~\ref{fig:entropy}(c), the three thin
curves show the loci of  $t_{\rm cool}=t_{\rm H}$ in the $T$-$z$ plane
for  $\delta=5$, 10,  and 20,  respectively.   Here we  have used  the
cooling  function  of   Sutherland  \&  Dopita  (1993)  with  a
metallicity  of  $0.01  Z_\odot$.   Effective radiative  cooling  only
occurs below these  curves.  Comparing these curves  with those
showing the temperature of the gas in pancakes, we see that heating by
gravitational pancaking at $z\la 2$  can have a
significant impact on  the subsequent accretion of gas into haloes that will be
low mass today.  At higher
redshifts,  however, the  cooling proceeds  sufficiently fast so that  the IGM
within the pancake can cool  back to its original temperature by the
time the low mass haloes within the pancake form.

Once again remember that the specific entropy generated by gravitational
pancaking at $z\sim 2$ is  very similar to what one needs to  suppress the 
cold gas fraction in low-mass haloes  (see \S\ref{sec:coldgas}).  Thus we 
conclude that the preheating  model discussed in the previous section,
which  is extremely successful  at explaining both the  stellar and
HI mass  functions, has a natural origin. One does not
need to invoke any star  formation or AGN activity; rather, the IGM
is preheated to the required specific entropy by the same process that
explains  the  formation of  the  dark  matter  haloes themselves.

\section{Discussion}
\label{sec:GF}

\subsection{Comparison with numerical simulations}
\label{ssec:simulations}

In the  previous sections we have  argued that
preheating  by gravitational pancaking causes
the  shapes of the HI  and stellar mass  functions at the  low mass
ends to agree with observations.
This begs the question as to why this effect  has not
already   been   seen   in   existing   hydrodynamical,   cosmological
simulations.  The short answer is  that no previous simulation had the
necessary resolution.  

To  study  the  effects  discussed  above
places  two different resolution constraints on  simulations.
First, they must be able to resolve
the shocks  that occur in the  forming pancakes and  second, they must
resolve the small galaxies that form within them.  The typical pancake
thickness  is  $\sim  200  h^{-1}\kpc$  in  comoving  units,  with  an
overdensity of  about 10.  In a Smoothed  Particle Hydrodynamics (SPH)
code, it requires  $4h_{\rm s}$ to resolve a  shock (Hernquist \& Katz 1989),
where $h_{\rm s}$ is  the SPH  smoothing  radius usually  chosen  so that there
are  32 particles within a  sphere of radius $2h_{\rm s}$.   Since the pancake
has a shock on both sides, the absolute minimum resolution has to be at
least $8h_{\rm s}$ across the  pancake thickness, i.e.  $h_{\rm s} \la
25  h^{-1}\kpc$.  Even  a shock  capturing Eulerian  grid or  AMR code
requires at least 4 grid cells across the pancake width to resolve the
post-shock structure.   The typical galaxy  that needs to  be affected
has a  dark halo  with a  circular velocity of  $\sim 50  \kms$, which
corresponds to a halo mass of $\sim 10^{10} \Msun$ and a virial radius
of $\sim 50 h^{-1} \kpc$.   To marginally resolve such haloes requires
at least 100  dark matter particles and hence a  particle mass for the
dark matter of less than $10^8  \Msun$.  In addition, one must have at
least 2 spatial resolution elements within the virial radius and hence
the  spatial resolution  must be  at least  $25 h^{-1}  \kpc$.   In an
Eulerian code both these spatial resolution requirements are identical
but in  a Lagrangian  code like SPH,  where the spatial  resolution is
variable,  the halo  requirement  is actually  $(200/10)^{1/3}=2.7$
times easier  to satisfy owing to  the higher overdensity. 

The  above  resolutions  have  not  been  achieved  in  any  published numerical
simulation.  The  highest resolution  simulation to
date at $z=0$ (Keres \etal 2005) has $128^3$ gas and dark matter particles in a
periodic, cubic volume  of $22.22 h^{-1}$ comoving Mpc  on a side.  At
an overdensity of 10 this simulation  has $h_{\rm s} = 81 h^{-1}$ kpc,
more than 3 times too large to properly resolve the pancake thickness.
In addition, the  dark matter particle mass of  $9\times 10^8\Msun$ is
almost an  order of magnitude  too large.  Furthermore, the  volume is
really too small to evolve the  simulation down to $z=0$ and to sample
enough different  environments.  Springel  \& Hernquist (2003)  have a
SPH simulation  in a large enough  volume, $100 h^{-1}$ Mpc  on a side
with $324^3$ gas and dark matter particles.  However, both the spatial
and mass resolution are worse than  in the case of Keres \etal (2005),
$143 h^{-1}$ kpc and $2\times 10^9 \msun$, respectively.  Their latest
unpublished  simulation  has $486^3$  particles  but  still has  worse
spatial  resolution ($96  h^{-1}$ kpc)  than Keres  \etal  (2005).  In
addition, the  particle mass  is $6\times  10^8 \Msun$, which is  still six
times too  large.  The  Eulerian simulation of  Kang \etal  (2005b) has
$1024^3$ cells in the same size volume as Springel \& Hernquist (2003)
for a grid cell size of  $97 h^{-1}$ kpc, again worse than Keres \etal
(2005).  Finally, Nagamine \etal  (2001) have  $768^3$ cells  within a
volume of $25 h^{-1}$ comoving Mpc on a side with a spatial resolution
of $75 h^{-1}$ kpc.

To perform a cosmological SPH  simulation in a uniform periodic volume
large enough to  evolve robustly to $z=0$, i.e. $100  h^{-1}$ Mpc on a
side,  and marginally  resolve  preheating, as  outlined above,  would
require  $1860^3$  dark  matter  and  gas  particles.   Such  a  large
simulation  is well  beyond the  reach  of the  current generation  of
computers and codes. However, one can reach such high resolutions in a
large volume by using a  ``zoom-in'' strategy, in which the resolution
is  only high  in  the  region of  interest  (Katz  \& White  1993).
Here, one starts with a large volume simulated at moderate resolution,
identify the  object whose  formation one wishes  to study,  trace the
particles  that end  up in  or near  this object  back to  the initial
conditions,  replace the particles  in this  Lagrangian region  with a
finer grid of  less massive particles, add the  small scale waves that
can  be  resolved  by  this  finer grid  to  the  initial  fluctuation
spectrum, and  re-run the simulation.  The particle  density away from
the  region of  interest is  reduced by  sparse sampling  the original
particle grid in a series of nested zones, always keeping the particle
density high  enough to maintain  an accurate representation  of tidal
forces.   This approach is  similar to  the one  used by  other groups
(e.g.   Navarro \& Steinmetz  1997, 2000;  Steinmetz \&  Navarro 1999,
Navarro \etal  1995; Sommer-Larsen  \etal 1999, Robertson  \etal 2004,
Governato \etal 2004) in their simulations of individual galaxies but one
needs to focus on present day low mass galaxies that form within pancakes at
higher redshift.
To  test   our   predictions  regarding   the  preheating   by
gravitational pancaking we plan to carry out investigations along this
line in the near future.

\subsection{Implications for galaxy formation and the IGM}

According to  the results  presented above, the  assembly of  gas into
galaxy-sized haloes is expected to proceed in two different modes with
a transition at $z\sim 2$. At $z>2$, the accreted intergalactic gas is
cold.   Since radiative  cooling is  efficient in  galaxy  haloes, gas
assembly into galaxies is expected to  be rapid and to be dominated by
clumps  of cold gas.   Combined with  the fact  that the  formation of
galaxy-sized  dark matter  haloes at  $z\ga 2$  is dominated  by major
mergers (e.g.  Li  \etal 2005), this suggests that  during this period
gas can  collapse into haloes  quickly to form starbursts  and perhaps
also feed  active galactic nuclei (e.g. Baugh et al. 2005).
Galactic feedback  associated with such
systems may  drive strong  winds into the  IGM, contaminating  the IGM
with metals.  Note  that strong feedback in this  phase is required to
reduce the  star formation efficiency in  high-$z$ galaxies; otherwise
too much gas would already turn  into stars at high $z$. This phase of
star formation  maybe  what is observed  as Lyman-break  galaxies and
sub-mm sources at  $z\ga 2$, and maybe  responsible for the formation
of elliptical  galaxies and the  bulges of spiral galaxies.   At $z\la
2$, however,  the situation is  quite different.  Since the  medium in
which galaxy-sized  haloes form is already  heated by previrialization
and since radiative  cooling is no longer efficient,  the accretion is
expected to be  dominated by hot, diffuse gas.   Such gentle accretion
of gas might be favorable for  the formation of the quiescent disks of
spiral galaxies.  Because  the accreted gas is diffuse  rather than in
cold clumps, it  can better retain its angular  momentum as it settles
into a  rotationally supported disk.  In addition,  since the baryonic
mass fraction  that forms the  disk is significantly smaller  than the
universal baryon fraction, the disk is less likely to become violently
unstable. Both effects may help alleviate the angular momentum problem
found  in some  numerical simulations,  i.e.  disks  that form  in CDM
haloes  have too  little  angular momentum  and  are too  concentrated
(Navarro \& Steinmetz 1997).
 
Depending on the halo  formation history, the bulge-to-disk ratio will
vary  from system  to system.   For haloes  that have  assembled large
amounts of mass before preheating,  the galaxies that form within them
should contain significant bulges, while in haloes that form after
preheating the galaxies should be dominated by a disk component. Since
haloes  that form later  are less  concentrated, this  model naturally
explains  why  later-type galaxies  usually  have  more slowly  rising
rotation curves.   An extreme example would  be low-surface brightness
(LSB) galaxies.   By definition, LSB  galaxies are disks in  which the
star formation efficiency  is low.  These galaxies are  also gas rich,
have high  specific angular momenta,  and show slowly  rising rotation
curves. The last two properties  are best explained if LSBs are hosted
by haloes that have formed only  in the recent past, since such haloes
have  low concentrations  (e.g.   Bullock \etal  2001; Wechsler  \etal
2002;  Zhao \etal  2003a,b), and  high spin  parameters  (e.g. Maller,
Dekel \& Somerville 2002; D'Onghia \& Burkert 2004;
Hetznecker \& Burkert 2005). However, there
is one  problem with such  a link between  LSBs and newly  formed dark
matter  haloes.  Since  the formation  of such  haloes  involves major
mergers,  these systems  are  expected to  produce strong  starbursts
rather than low-surface brightness disks.  This problem does not exist
in our model since these haloes are expected to  form in a preheated
medium and gas  accretion into such haloes will be smooth. 
It  will be  interesting to  see if  our model  can predict  the right
number of LSB  galaxies and explain the existence  of extreme systems
like Malin I.

Our model also  opens a new avenue to understand  the evolution of the
galaxy population with redshift. As we argued above, star formation at
$z>2$ is expected to be  dominated by starbursts associated with major
mergers of gas rich systems, while star formation at $z<2$ is expected
to occur  mostly in quiescent  disks. This has  important implications
for understanding the  star formation history of the  universe and for
understanding the  evolution of the galaxy population  in general. For
example, our model implies a characteristic redshift, $z\sim 2$, where
both  the  star  formation   history  and  galaxy  population  make  a
transition from a  starburst-dominated mode to a more  quiescent mode. 
Furthermore,  if  AGNs  are   driven  by  gas-rich  major  mergers,  a
transition at $z\sim 2$ is  also expected in the AGN population. There
are some hints  about such transitions in the  observed star formation
history (e.g. Blain \etal 1999), and in the observed number density of
AGNs (e.g.  Shaver \etal 1996).   Recent observations of  damped Lyman
alpha systems also suggests a change  in behavior at $z\sim 2$ both in
the cold  gas content and in  the number density of  such systems (see
Rao 2005).

 The preheated medium  we envision here is closely related to
the  warm-hot intergalactic  medium (WHIGM)  under intensive  study in
recent years.  Hydrodynamical simulations  show that between 30 and 40
percent  of all baryons  reside in  this WHIGM,  which is  produced by
shocks  associated   with  gravitational  collapse   of  pancakes  and
filaments  (e.g. Cen  \& Ostriker  1999; Dave  \etal 2001;  Kang \etal
2005b).   These  results are  consistent  with observational  estimates
based  on  the  study  of  UV  absorption  lines  in  the  spectra  of
low-redshifts QSOs (see Tripp \etal  2004 for a review).  In
our  model,   the  low-temperature  component  of   this  WHIGM, i.e, with
temperatures of a  few times $10^5$K, is associated  with pancakes of
relatively low mass ($\sim 5  \times 10^{12} h^{-1} \Msun$), within
which late-type galaxies form. As we discussed in \S\ref{ssec:simulations},
current simulations are still unable to
make accurate  predictions regarding this particular  component of the
WHIGM.   Future simulations  of higher  resolution, however,  may shed
light on the relation between the properties of galaxies and
those of the IGM in  their immediate surroundings.  Such a relationship
may prove  pivotal in observational  hunts for the  missing baryons
as  the spatial  distribution of galaxies and their  properties can serve
as guideposts in the search for the WHIGM.

\section{Conclusions}
\label{sec:concl}

Understanding  the shallow  faint-end slope  of the  galaxy luminosity
function, or equivalently the stellar  mass function, is a well known
problem in galaxy formation. In the standard model it is often assumed
that supernova feedback keeps large fractions of the baryonic material
hot,  thus suppressing  the  amount of  star  formation. Although  the
efficiency of this process might be tuned so that one fits the faint-end
slope of the  galaxy luminosity function it has a  number of problems. 
First, the required efficiencies  are extremely high and are
inconsistent  with detailed  numerical simulations  (e.g.,  Mac-Low \&
Ferrara 1999). Second,  unless the hot gas is  somehow expelled from
the  dark  matter  halo  forever, which  requires  even  higher
feedback  efficiencies,  this  model  predicts  hot  haloes  of  X-ray
emitting  gas  around  disk   galaxies,  which  is  inconsistent  with
observations  (e.g.,  Benson  \etal  2000).   In this  paper  we  have
demonstrated  that this  model has  an additional  shortcoming.  Using
recently obtained HI  mass functions we show  that the supernova
feedback model  predicts HI  masses that are  too high. The  reason is
that supernova feedback requires star formation which in turn requires
high surface densities  of cold gas.  The latter  owes to the existence
of a star  formation threshold density, which has  strong support from
both  theory (e.g.,  Quirk 1972,  Silk 2001)  and  observations (e.g.,
Kennicutt 1989; Martin \& Kennicutt 2001).

We therefore argue that  simultaneously matching the shallow, low-mass
slopes of  the stellar and  HI mass functions requires  an alternative
mechanism, which does  not directly depend on star  formation. We
demonstrate that  a mechanism that can  preheat the  IGM to a
specific gas entropy of $\sim 10\,{\rm keV\,cm^{2}}$, can fit both the
observed stellar mass function as well as the HI mass function. We
also show that gravitational  instability of the cosmic density field
can  be the  source  of this  preheating.  Our  idea is
fairly  simple:   low  mass   haloes  form   within  larger-scale
overdensities that  have already undergone collapse  along their first
axis.   This  `pancake' formation  causes  the  associated  gas to  be
shock-heated, and  providing that the gas cooling rate is  sufficiently slow,
the low  mass haloes embedded within these  pancakes subsequently form
in a preheated medium.

Using the  ellipsoidal collapse model,  we demonstrate  that the
progenitors    of    present-day     haloes    with    masses    $M\la
10^{12}h^{-1}\msun$ were embedded in  pancakes of masses $\sim 5\times
10^{12}h^{-1}\msun$ at  $z\sim 2$. The formation of  such pancakes can
heat the  gas associated with them  to a temperature  of $\sim 5\times
10^5{\rm K}$ and compress it to  an overdensity of $\sim 10$. This gas
has a cooling time longer than  the age of the Universe and a specific
entropy  of  about  $15{\rm  keV}\cm^2$;  the  amount
needed to explain the observed stellar and HI mass functions.

Our results demonstrate  that heating associated with previrialization
may also help  solve a  number of outstanding problems  in galaxy
formation  within  a  CDM  universe. However, detailed,  high-resolution
numerical simulations  will be
required to  test  our  predictions  in detail.
Such simulations will help us understand how the formation of a pancake
heats the gas initially associated with it and whether the amount of heating
follows our analytic results. For example, our 
calculation indicates that shock heating will dominate over cooling in
typical pancakes at $z \la 2$. However, such calculations assume that
the gas is smoothly distributed.  Structures at scales smaller than the pancake
could lead to density inhomogeneities, either pancakes, filaments or halos,
and these could promote extra cooling and move the transition to lower
redshifts.  A similar effect occurs when one calculates the transition mass
from cooling dominated to infall dominated accretion in galaxy formation
(White \& Frenk 1992) and compares it with actual simulations
(Keres et al 2005). They will also help us understand 
how such heating affects subsequent gas accretion and cooling
into the dark haloes that form in the pancake and hopefully
derive actual stellar and HI mass functions at the small mass end.
We have  argue that  no  cosmological, hydrodynamical
simulation to date has the  required spatial and/or mass resolution to
study the  pancake preheating proposed  here. To  achieve the required
numerical  resolution, we  suggest  resimulating, at  high
resolution, a  number of pancakes  (and filaments) with masses  of the
order  of  $5  \times  10^{12}  h^{-1}  \Msun$,  identified  from  large
cosmological,  hydrodynamical simulations.   Thus  far this  `zoom-in'
strategy has mainly  been used to study clusters  and galaxies. If our
estimates are correct, it will  be extremely interesting to apply this
technique to study pancakes and their enclosed, low mass haloes.


\section*{Acknowledgement}

We  thank Jessica  Rosenberg and  Martin Zwaan  for providing  us with
their HI  mass functions.   FvdB is grateful  to Aaron  Dutton, Justin
Read and Greg Stinson for valuable discussion.
NSK was supported by NSF AST-0205969, NASA NAGS-13308, and 
NASA NNG04GK68G.


\end{document}